\documentclass{sig-alternate}

\usepackage{graphicx}% Include figure files
\usepackage{dcolumn}% Align table columns on decimal point
\usepackage{bm}% bold math

\begin{document}

\title{Networking Behavior in Thin Film and Nanostructure Growth Dynamics}

\numberofauthors{3}
\author{
% 1st. author
\alignauthor Murat Yuksel \\
       \affaddr{University of Nevada - Reno}\\
%       \affaddr{Computer Science and Engineering Department}\\
       \affaddr{Reno, NV 89557, USA}\\
       \email{yuksem@cse.unr.edu}
% 2nd. author
\alignauthor Tansel Karabacak\\
       \affaddr{University of Arkansas}\\ % at Little Rock}\\
%       \affaddr{Applied Science Department}\\
       \affaddr{Little Rock, AR 72204, USA.}\\
       \email{txkarabacak@ualr.edu}
% 3rd. author
\alignauthor Hasan Guclu\\
       \affaddr{Los Alamos National Lab}\\
%       \affaddr{Center for Nonlinear Studies, Theoretical Division}\\
       \affaddr{Los Alamos, NM 87545, USA.}\\
       \email{guclu@lanl.gov}
%\and  % use '\and' if you need 'another row' of author names
}

%\date{\today}

\maketitle

\begin{abstract}
\vspace{-2mm} Thin film coatings have been essential in development
of several micro and nano-scale devices. To realize thin film
coatings various deposition techniques are employed, each yielding
surface morphologies with different characteristics of interest.
Therefore, understanding and control of the surface growth is of
great interest. In this paper, we devise a novel network-based
modeling of the growth dynamics of such thin films and
nanostructures. We specifically map dynamic steps taking place
during the growth to components (e.g., nodes, links) of a
corresponding network. We present initial results showing that this
network-based modeling approach to the growth dynamics can simplify
our understanding of the fundamental physical dynamics such as
shadowing and re-emission effects.
\end{abstract}

\vspace{-3mm}


%\keywords{Nano-scale growth, dynamic network models} % NOT required for Proceedings

\vspace{-3mm}
\section{Introduction}
\vspace{-1mm}

Thin film coatings have been the essential components of various
devices in industries including microelectronics, optoelectronics,
detectors, sensors, micro-electro-mechanical systems (MEMS), and
more recently nano-electro-mechanical systems (NEMS). These coatings
have thicknesses typically in the nano- to micro-scales and are
grown using vacuum deposition techniques
%\cite{SMITH95,MEAKIN98,
\cite{BARABASI95}. Thin film surface morphology controls many
important physical and chemical properties of the films. It is
therefore of great interest to understand and control the
development of the surface morphology during thin film growth.

Commonly employed deposition techniques are \emph{thermal
evaporation, sputter deposition, chemical vapor deposition (CVD)},
and \emph{oblique angle deposition}. Different than others, oblique
angle deposition technique \cite{KARABACAK05}
%,KARABACAK04 10,LAKHTAKIA05 12}
is typically used for the growth of nanostructured arrays of rods
and springs through a physical self-assembly process. In many
applications, it is often desired to have atomically flat thin film
surfaces. However, in almost all of the deposition techniques
mentioned above, the surface morphology generates a growth front
roughness. The formation of growth front is a complex phenomenon and
very often occurs far from equilibrium. When atoms are deposited on
a surface, atoms do not arrive at the surface at the same time
uniformly across the surface. This random fluctuation, or noise,
which is inherent in the process, may create the surface roughness.
The noise competes with surface smoothening processes, such as
surface diffusion (hopping), to form a rough morphology if the
experiment is performed at either a sufficiently low temperature or
a high growth rate.

A conventional statistical mechanics treatment cannot be used to
describe this complex phenomenon. About two decades ago, a dynamic
scaling approach \cite{FAMILY85,FAMILY86} was pro posed to describe
the morphological evolution of a growth front. Since then, numerous
modeling and experimental works have been reported based on this
dynamic scaling analysis \cite{BARABASI95}. On the other hand, there
has been a significant discrepancy among the predictions of these
growth models and the experimental results published \cite{LU03}.
Briefly, theoretical predictions of growth models in dynamic scaling
theory basically fall into two categories. One involves various
surface smoothing effects, such as surface diffusion. The other
category involves the shadowing effect (which originates from the
preferential deposition of obliquely incident atoms on higher
surface points and always occurs in sputtering and CVD) during
growth. However, experimentally reported values of growth exponent
(which measures how fast the root-mean-square roughness of the
surface evolves as a function of time according to a power-law
relation) are far from agreement with the predictions of these
growth models. Especially, sputtering and CVD techniques are
observed to produce morphologies ranging from very small to very
large growth exponent values.

Understanding the thin film and nanostructure growth dynamics under
the above-mentioned deposition techniques has been of high
importance. There have been several studies revealing fundamental
dynamic effects (e.g., \emph{shadowing, re-emission,
surface-diffusion, and noise effects}) taking place during the
growth process. Studies towards explaining the growth dynamics have
been partly successful and only the simulation-based studies were
able to include all these effects. In this paper, we devise a novel
network-based modeling approach to better understand the growth
dynamics. We define a concise mapping between a network and the
basic physical operations taking place in the growth process. We,
then, develop qualitative and quantitative understanding of the
growth dynamics by studying the corresponding network model. We
present our initial results based on previously recorded simulations
of the growth process.

The rest of the paper is organized as follows: We start with
covering the thin film and nanostructure growth process and the
basic physical effects involved in
Section~\ref{sec:basics-of-growth}. We then survey the applications
of dynamic network models on various areas in
Section~\ref{sec:network-models}. Section~\ref{sec:mapping}
describes the details of our methodology of mapping growth dynamics
to a network. We present initial results of our network-based
modeling approach in Section~\ref{sec:results}, and conclude in
Section~\ref{sec:conclusion}.

\vspace{-3mm}
\section{Basics of Thin Film and Nanostructure Growth}
\label{sec:basics-of-growth} \vspace{-1mm}

\begin{figure}
\includegraphics[keepaspectratio=true,angle=0,width=85mm]{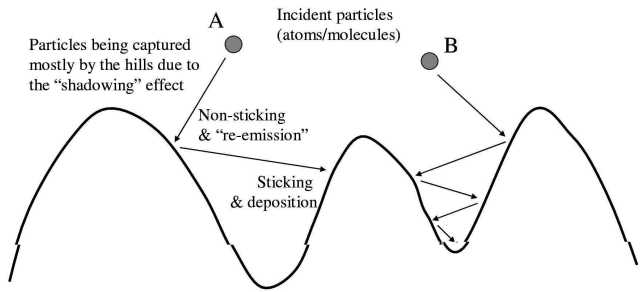}
\vspace{-8mm} \caption{Surface of a growing thin film (growth front)
under shadowing and re-emission effects.} \label{fig:fig_3}
\vspace{-6mm}
\end{figure}

Only recently, it has been recognized that in order to better
explain the dynamics of surface growth one should take into account
the effects of both ``shadowing'' and ``re-emission'' processes
\cite{KARABACAK01}. As illustrated in Figure \ref{fig:fig_3},
particles can approach the surface at oblique angles and be captured
by higher surface points (hills) due to the shadowing effect. This
leads to the formation of rougher surfaces with columnar structures
that can also be engineered to form ``nanostructures'' under extreme
shadowing conditions, as in the case of oblique angle deposition
that can produce arrays of nanorods and nanosprings
\cite{KARABACAK05}.
%,KARABACAK04,LAKHTAKIA05}.
In addition, depending on the detailed deposition process, particles
can either stick to or bounce off from their impact points, which is
determined by a sticking probability, also named ``sticking
coefficient'' (s). Non-sticking particles are re-emitted and can
arrive at other surface points including shadowed valleys. In other
words, re-emission has a smoothening effect while shadowing tries to
roughen the surface. Both the shadowing and re-emission effects have
been proven to be dominant over the surface diffusion and noise, and
act as the main drivers of the dynamical surface growth front
\cite{KARABACAK05}. The prevailing effects of shadowing and
re-emission rely on their ``non-local'' character: The growth of a
given surface point depends on the heights of near and far-away
surface locations due to shadowing and existence of re-emitted
particles that can travel over long distances.

Due to the complexity of the shadowing and re-emission effects, no
growth model has been developed yet within the framework of
dynamical scaling theory that take into both these effects and still
that can be analytically solved to predict the morphological
evolution of thin film or nanostructure deposition. A dynamic growth
equation that was proposed by Drotar et al. \cite{DROTAR00} and
developed for plasma and reactive ion etching processes (where in
etching surface atoms are removed instead of being incorporated to
the surface as in the case of deposition) that take into the
re-emission and shadowing effects could only be solved numerically
for a limited case of re-emission and shadowing scenarios.  Only
recently, shadowing and re-emission effects could be fully
incorporated into the Monte Carlo lattice simulation approaches
\cite{LU03,KARABACAK01,DROTAR00}.
%These simulations successively predicted the experimental results
%including the $\beta$ values shown in Figure \ref{fig:fig_2}.

%Moreover, it has been very recently revealed that shadowing effect
%can lead to the breakdown of dynamical scaling theory due the
%formation of a mounded surface morphology
%\cite{PELLICCIONE06,PELLICCIONE06_2}. In these studies, using Monte
%Carlo simulations it has been shown that for common thin film
%deposition techniques, such as sputter deposition and CVD, a
%``mound'' structure can be formed with a characteristic length scale
%that describes the separation of the mounds, or ``wavelength''
%$\lambda$ . It has been found that the temporal evolution of
%$\lambda$ is distinctly different from that of the mound size, or
%the lateral correlation length, $\xi$. The formation of the mound
%structure is due to non-local growth effects, such as shadowing,
%that lead to the breakdown of the self-affinity of the morphology
%described by the dynamic scaling theory. The wavelength grows as a
%function of time in a power law form, $\lambda \sim t^p$, where
%$p\simeq 0.5$  for a wide range of growth conditions, while the
%mound size grows as $\xi \sim t^{1/z}$ , where $1/z$ depends on the
%growth conditions.

In brief, conventional growth models, which do not include
re-emission effects, in dynamic scaling theory can not explain most
of the experimental results reported for dynamic thin film growth.
On the other hand, simulation techniques that include re-emission
effects along with other important processes such as shadowing,
surface diffusion, and noise can successively predict the
experimental results but can not always be easily implemented by a
widespread of researchers.

\vspace{-3mm}
\section{Dynamic Network Models}
\label{sec:network-models} \vspace{-1mm}

The study of complex networks pervades various areas of science
ranging from sociology to statistical physics \cite{BOCA06}.
%,NEWMAN03,DORO02,ALBERT02}.
A network in terms of modeling can be defined as a set nodes with
links connecting them. Examples of real life complex networks
include the Internet, the World Wide Web, metabolic networks,
transportation networks, social networks, etc.
%Regular lattices are commonly used to study physical systems with
%short-range or long-range interactions. Earlier network studies
%focused mostly on the topological properties of the networks.
Recent works, motivated by a large number of natural and artificial
systems, such as the ones listed above, have turned the focus onto
processes on networks, where the interaction and dynamics between
the nodes are facilitated by a complex network.
%The question then is
%how the collective behavior of the system is influenced by this
%possibly complex interaction topology.
Here, our aim is to construct the network from the apparent
dynamics. These systems also typically constitute large scale
elements unlike the atomic processes involved during thin film or
nanostructure growth.

By using network-based modeling, fundamental understanding of many
natural and artificial systems has been attained. In complex
networks research, two major types of network models are used for
various applications: Small-world \cite{WATTS98} and scale-free
(power-law) \cite{BARABASI99} networks. Watts and Strogatz, inspired
by a sociological experiment, have proposed a network model known as
the small-world (SW) network, which means that, despite their often
large size, there is a relatively short path between any two nodes
in most networks with some degree of randomness. The SW network was
originally constructed as a model to interpolate between regular
lattices and completely random networks. Systems and models (with
well known behaviors on regular lattices) have been studied on SW
networks, such as the Ising model, phase ordering, the
Edwards-Wilkinson model, diffusion, and resistor networks.

The other major type of network is based on an observation made in
the context of real networks such as the Internet, World Wide Web,
scientific collaboration network, and e-mail network. The common
characteristic among these networks is that they all exhibit
power-law degree (connectivity) distributions. These networks are
commonly known as power-law or scale-free networks \cite{BARABASI99}
since their degree distributions are free of scale (i.e., not a
function of the number of nodes $N$) and follow power-law
distributions over many orders of magnitude. This phenomenon has
been represented by the probability of having nodes with k degrees
as $P(k) \sim k^{-\gamma}$ where $\gamma$ is usually between 2 and
3. The origin of the scale-free behavior can be traced back to two
mechanisms that are present in many systems, and have a strong
impact on the final topology. First, networks are developed by the
addition of new nodes that are connected to those already present in
the system. This mechanism signifies continuous expansion in real
networks. Second, there is a higher probability that a new node is
linked to a node that already has a large number of connections.
With appropriate mapping to a network model, both of these
mechanisms can be qualitatively shown in thin film and nanostructure
growth dynamics. If we consider the thin film surface as a set of
nodes and re-emissions as the links between them, the first
mechanism refers to the understanding that each particle gets
``connected'' to the grid network by falling on to the film surface.
Similarly, the second mechanism refers to that a falling particle
will more likely to land on a large-size node thereby contributing
to the scale-free topological behavior of the growth dynamics.

\begin{figure}
\includegraphics[keepaspectratio=true,angle=0,width=85mm]{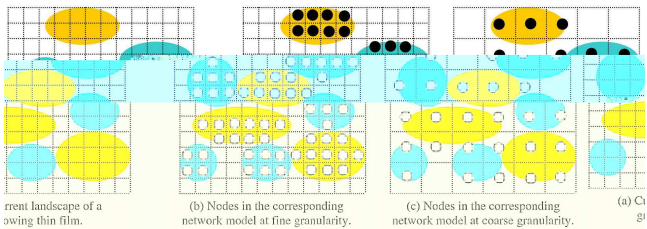}
\vspace{-8mm} \caption{Identification of network ``nodes'' in a grid
network model corresponding to a landscape of a growing thin film.
%Resolution of the grid can have significant impact on the number of
%nodes.
% in the network model.
%, yielding potential un-captured growth dynamics if the resolution
%is too coarse as in (c).
} \label{fig:fig_4} \vspace{-3mm}
\end{figure}

\begin{figure}
\includegraphics[keepaspectratio=true,angle=0,width=85mm]{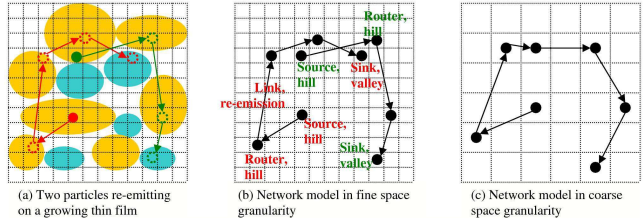}
\vspace{-8mm} \caption{Grid network model development in space:
Consider two, red and green, particles falling on a growing thin
film sample. The red particle makes four re-emissions while the
green one makes three re-emissions. We model each re-emission as a
``link'' between the nodes corresponding to the starting and ending
points of the re-emission.
%The corresponding nodes to these starting and ending points can be
%aggregated over space as in (c), resulting in a different
%topological behavior.
} \label{fig:fig_5} \vspace{-6mm}
\end{figure}

\vspace{-3mm}
\section{Mapping Growth Dynamics to a Network Model}
\label{sec:mapping} \vspace{-1mm}

Interestingly, non-local interactions among the surface points of a
growing thin film that originate from shadowing and re-emission
effects can lead to non-random preferred trajectories of
atoms/molecules before they finally stick and get deposited. For
example, during re-emission, the path between two surface points
where a particle bounces off from the first and head on to the
second can define a ``network link'' between the two points. If the
sticking coefficient is small, then the particle can go through
multiple re-emissions that form links among many more other surface
points. In addition, due to the shadowing effect, higher surface
points act as the locations of first-capture and centers for
re-emitting the particles to other places. In this manner, hills on
a growing film resembles to the network ``nodes'' of heavy traffic,
where the traffic is composed by the amount of re-emitted particles.

\begin{figure}
\includegraphics[keepaspectratio=true,angle=0,width=85mm]{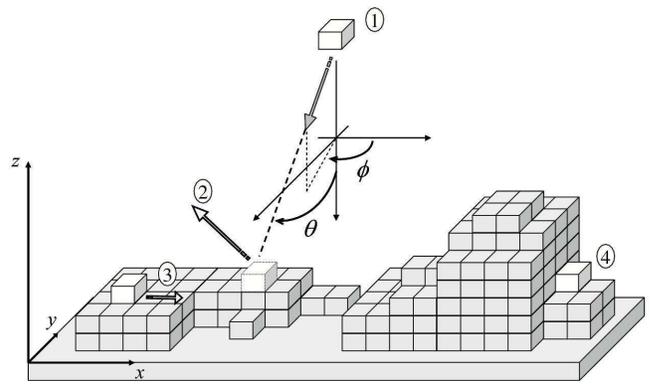}
\vspace{-8mm} \caption{Some basic processes in the simulation: (1) A
particle is sent towards surface with angles $\theta$ and $\phi$
based on an angular distribution chosen based on the deposition
technique. This particle sticks to the surface with probability
$s_0$. (2) If it does not stick, then it is re-emitted after which
it may find another surface feature and stick there with probability
$s_1$. This re-emission process continues like this for higher-order
particles, too. (3) An adatom can diffuse on the surface. (4) Some
surface points are shadowed from the incident and re-emission fluxes
of particles due to the nearby higher surface features.
%In all these processes, trajectories of particles can be tracked in
%order to reveal the dynamic network behavior in detail.
} \label{fig:fig_6} \vspace{-6mm}
\end{figure}

Several issues need to be considered in making a useful and
appropriate mapping between the growth dynamics of thin films and
nanostructures to a network modeling framework. Let us consider a
snapshot of a growing thin film's landscape. In Figure
\ref{fig:fig_4}(a), let us say that blue color shows currently
elevated (i.e. hills) regions of the film and yellow color shows
currently not elevated (i.e. valleys) regions of the film. The first
mapping issue is to define a ``node'' in the corresponding network
model. That is, what should be the boundary of the corresponding
network node on the thin film surface? Intuitively, each blue or
yellow region in Figure \ref{fig:fig_4}(a) should ideally get mapped
to a network node. However, this depends on the resolution of the
grid being used for developing a network model. If the grid
resolution is too fine, then a blue/yellow region of the film can
correspond to multiple nodes as in Figure \ref{fig:fig_4}(b).
Conversely, if the grid resolution is too coarse, then multiple
blue/yellow regions can correspond to one network node as in Figure
\ref{fig:fig_4}(c). Having finer grid is more likely to capture
dynamics of the growth; therefore, we will develop our network
models in as fine granularity as possible. For a fine granularity
%(i.e. multiple blue/yellow regions corresponding to a node)
network model, it is always possible to aggregate the data
pertaining to neighboring nodes and observe the behavior at coarser
granularity. This is illustrated in Figure \ref{fig:fig_5}, where
the grid network model can be developed at various scales in space.

After fixing the placement of nodes on the thin film, we then map
growth dynamics to components of the corresponding grid network
model as shown in Figure \ref{fig:fig_5}. In general, we argue that
we can make an analogy that hills and valleys are nodes of the
network system, but hills act as distributing centers, and valleys
as gathering centers due to the shadowing and re-emission effects,
respectively. The re-emissions of particles can, then, be modeled as
a ``link'' from the re-emission's starting node to the re-emission's
ending node. The time it takes for the particle to reach to its new
point can be considered as the link's ``propagation delay'', which
implicitly expresses the distance between the starting and the
ending nodes of the re-emission. It is even possible to consider the
link's ``capacity'' as the highest possible number of particles that
can simultaneously travel from the starting and the ending nodes of
the re-emission, which is limited by the physical space
corresponding to the link and average size of the re-emitting
particles.

%In terms of network traffic, nodes can be classified as: source,
%sink, or router. So, the initial point where an atom re-emits can
%correspond to a ``source'' in a network, and the final point where
%the atom sticks/settles can be thought as a ``sink'' in the network.
%Similarly, the intermediate re-emission points can be thought as the
%``routers'' as illustrated in Figure \ref{fig:fig_5}. Further, a
%``traffic model'' for the thin film growth can then be constructed
%by counting the number of atoms starting from a source and sticking
%at a sink on the film.
%%That is, if we consider the film as a grid with two dimensions, the
%%traffic from point (i,j) to the point (u,v) can be measured by the
%%amount of atoms initially falling at (i,j) but ending up at (u,v)
%%after several reemissions.
%This traffic can be modeled at various time-scales depending on the
%total growth time of the film, so that one can talk about ``atoms
%per second'' as the traffic unit. This approach can reveal the
%effect of one surface point on the other points, which can be
%modeled by the ``routing'' phenomenon of networks. Just like the
%network model development over space is shown in Figure
%\ref{fig:fig_5}(b)-(c), it is possible to construct different
%network models in time for the same thin film growth dynamics at
%varying time-scales. This shows that network-based modeling of thin
%film growth is very flexible and can capture all possible aspects of
%the dynamics involved.

\begin{figure}
\includegraphics[keepaspectratio=true,angle=0,width=85mm]{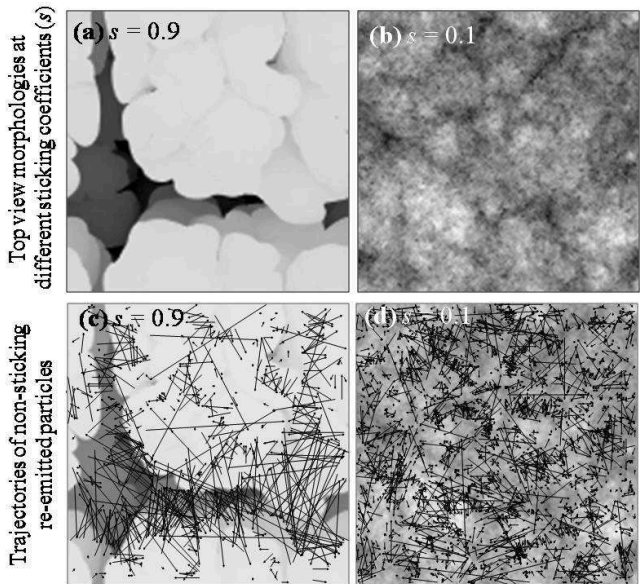}
\vspace{-8mm} \caption{Top view images of simulated thin film
surfaces grown under shadowing, re-emission, and noise effects
%(no surface diffusion is included in these simulations)
for sticking coefficients (a) $s = 0.9$ and (b) $s = 0.1$.
%The incident flux of particles has an angular distribution designed
%for chemical vapor deposition (CVD).
Corresponding projected trajectories of the re-emitted particles are
also mapped on the top view morphologies for (c) $s = 0.9$ and (d)
$s = 0.1$.
%Qualitative network behavior can be seen among surface points linked
%by the re-emission trajectories.
} \label{fig:fig_7} \vspace{-6mm}
\end{figure}

Since it is not possible to experimentally track the trajectories of
re-emitted and deposited atoms during dynamic thin film growth, we
will use Monte Carlo simulation approaches instead that were already
shown to efficiently mimic the experimental processes and correctly
predict the dynamic growth morphology. In these simulations, each
incident particle (e.g., atom or molecule) is represented with the
dimension of one lattice point. A specific angular distribution for
the incident flux of particles is chosen depending on the deposition
technique being simulated. At each simulation step, a particle is
sent toward a randomly chosen lattice point on the substrate
surface. Depending on the value of sticking coefficient $s$, the
particle can bounce off and re-emit to other surface points. At each
impact sticking coefficient can have different values represented as
$s_n$, where $n$ is the order of re-emission ($n=0$ being for the
first impact)\footnote{In this paper we assume a constant sticking
coefficient for all subsequent re-emissions.}. In all the emission
and re-emission processes shadowing effect is included, where the
particle's trajectory can be cut-off by long surface features on its
way to other surface points. After the incident particle is
deposited onto the surface, it becomes a so called ``adatom''.
Adatoms can hop on the surface according to some rules of energy,
which is a process mimicking the surface diffusion. This simulation
steps are repeated for other particles being sent onto the surface.
Figure \ref{fig:fig_6} illustrates the basic growth processes
included in a typical Monte Carlo simulation approach.
%In Monte Carlo simulations, trajectories of particles
%during deposition, re-emission, and surface diffusion can be tracked
%in order to reveal the dynamic network behavior in detail.

\begin{figure}
\includegraphics[keepaspectratio=true,angle=0,width=85mm]{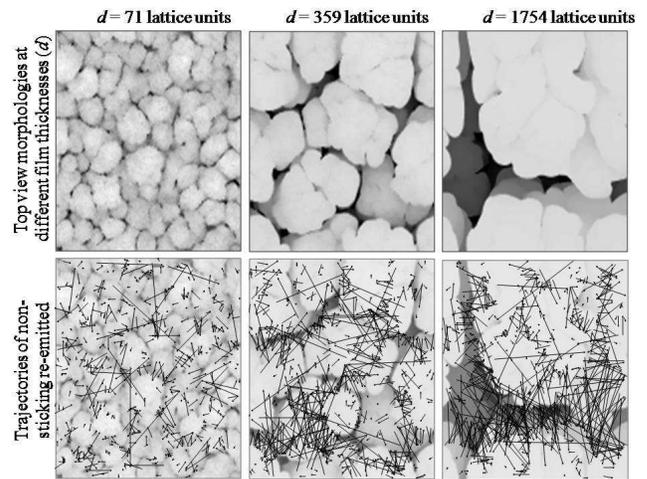}
\vspace{-8mm} \caption{First row: Top view images from the simulated
thin film surfaces for a CVD growth with $s = 0.9$ at different film
thicknesses $d$. Bottom row: Corresponding projected trajectories of
the re-emitted particles qualitatively show the dynamic change in
the network topography. } \label{fig:fig_8} \vspace{-6mm}
\end{figure}

\vspace{-3mm}
\section{Initial Results} \label{sec:results}
\vspace{-1mm}

\begin{figure*}
\begin{center}
\begin{tabular}{cccc}
\hspace{-5mm}\includegraphics[keepaspectratio=true,width=45mm]{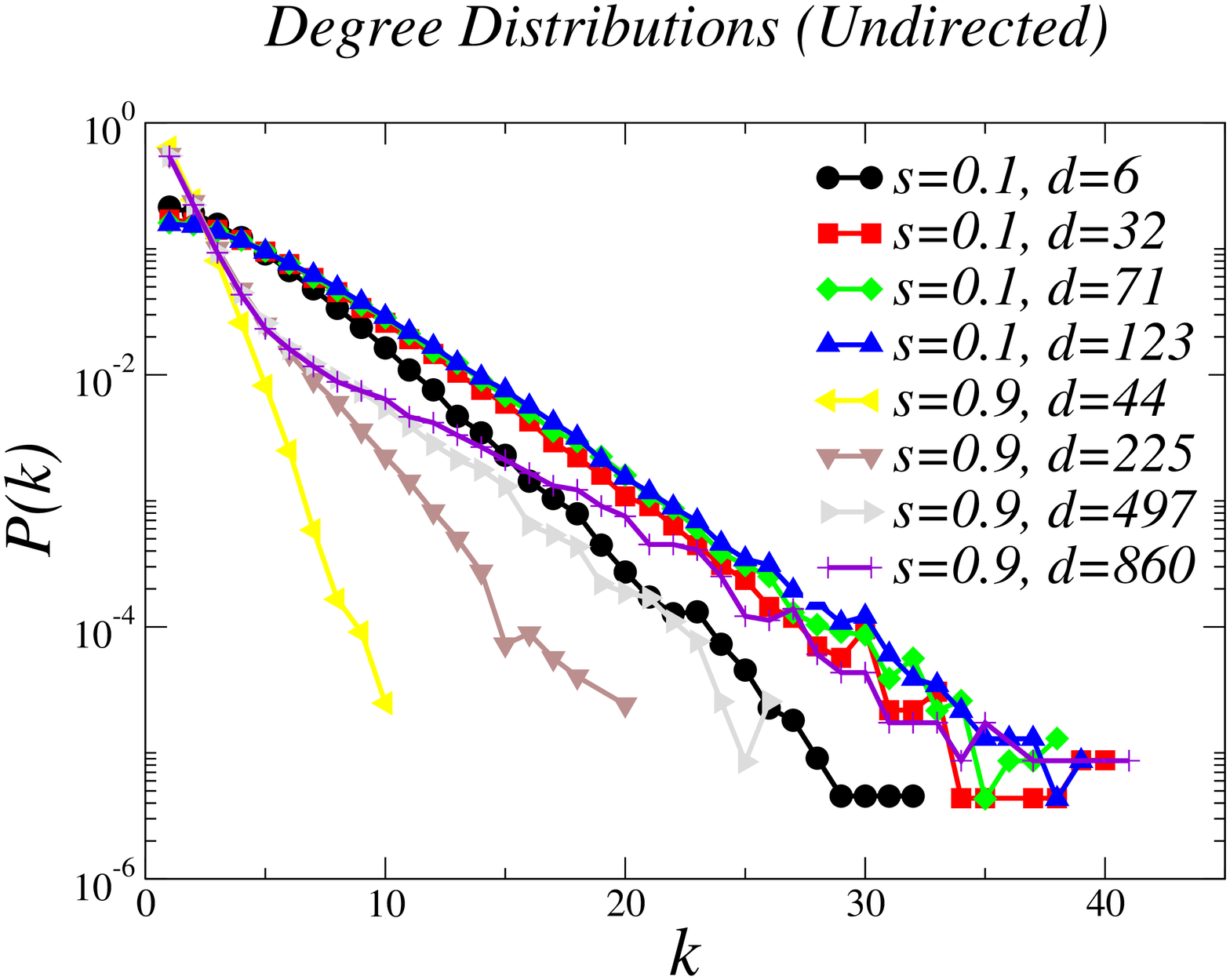}
&
\hspace{-3mm}\includegraphics[keepaspectratio=true,width=45mm]{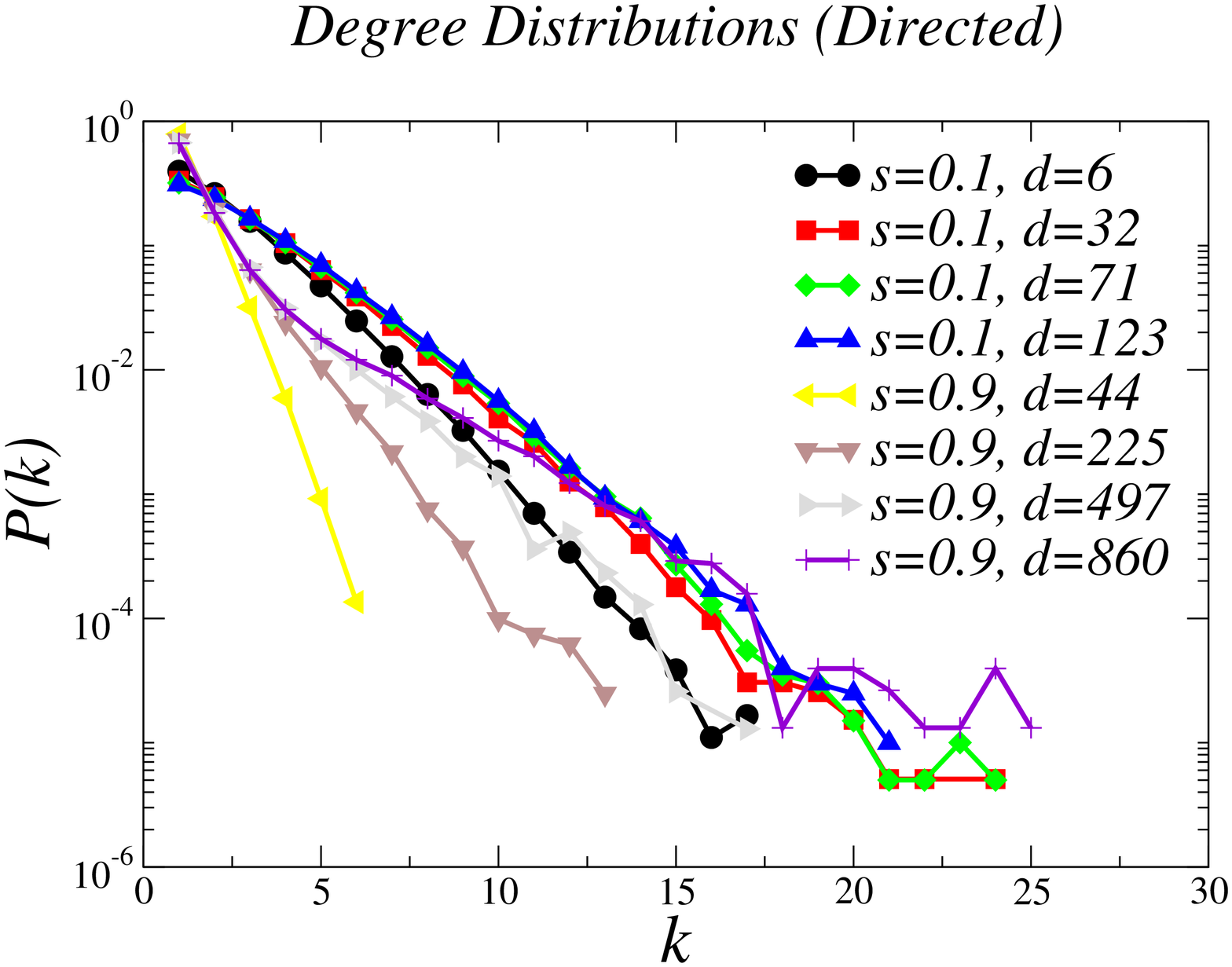}
&
\hspace{-3mm}\includegraphics[keepaspectratio=true,width=45mm]{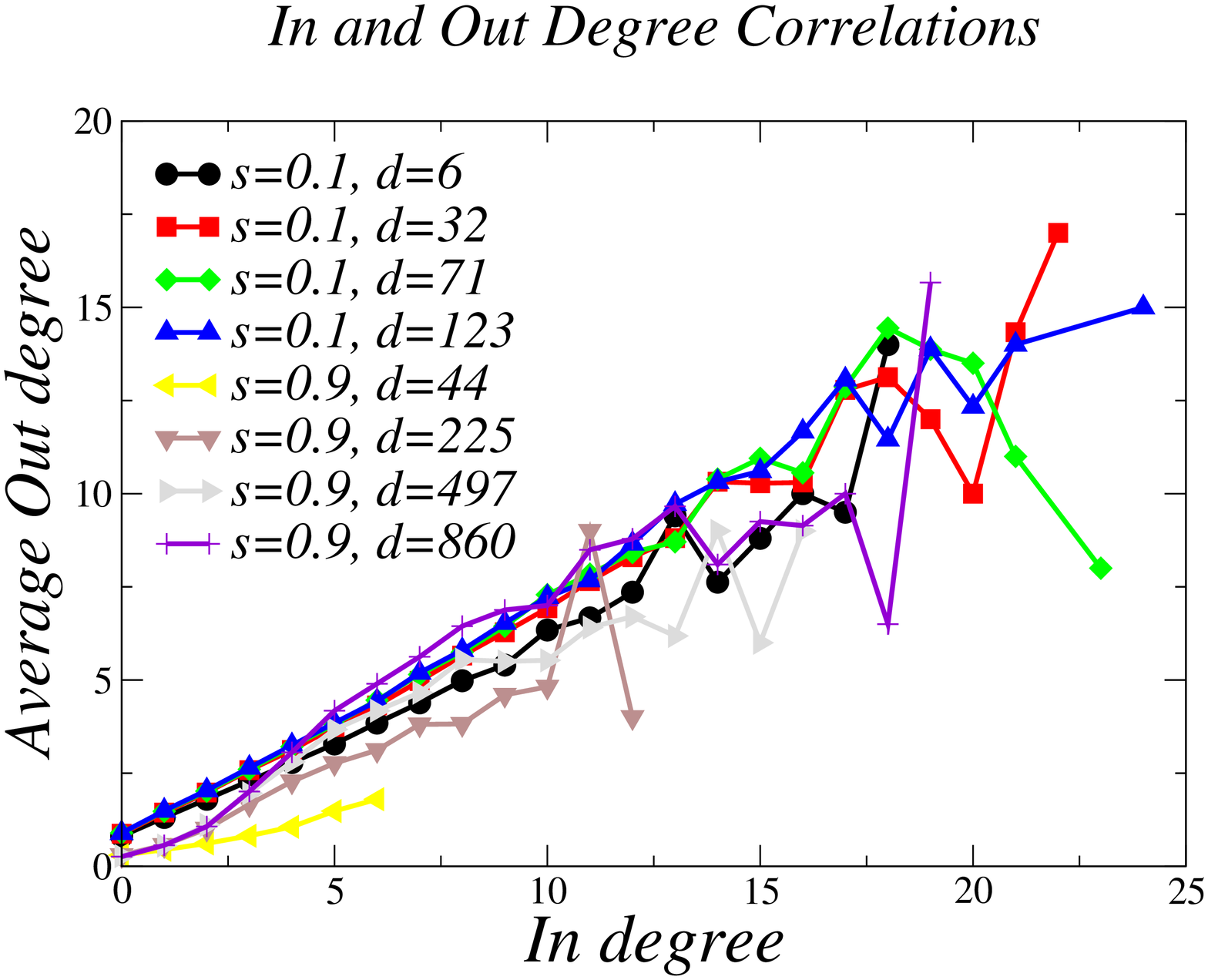}
&
\hspace{-3mm}\includegraphics[keepaspectratio=true,width=45mm]{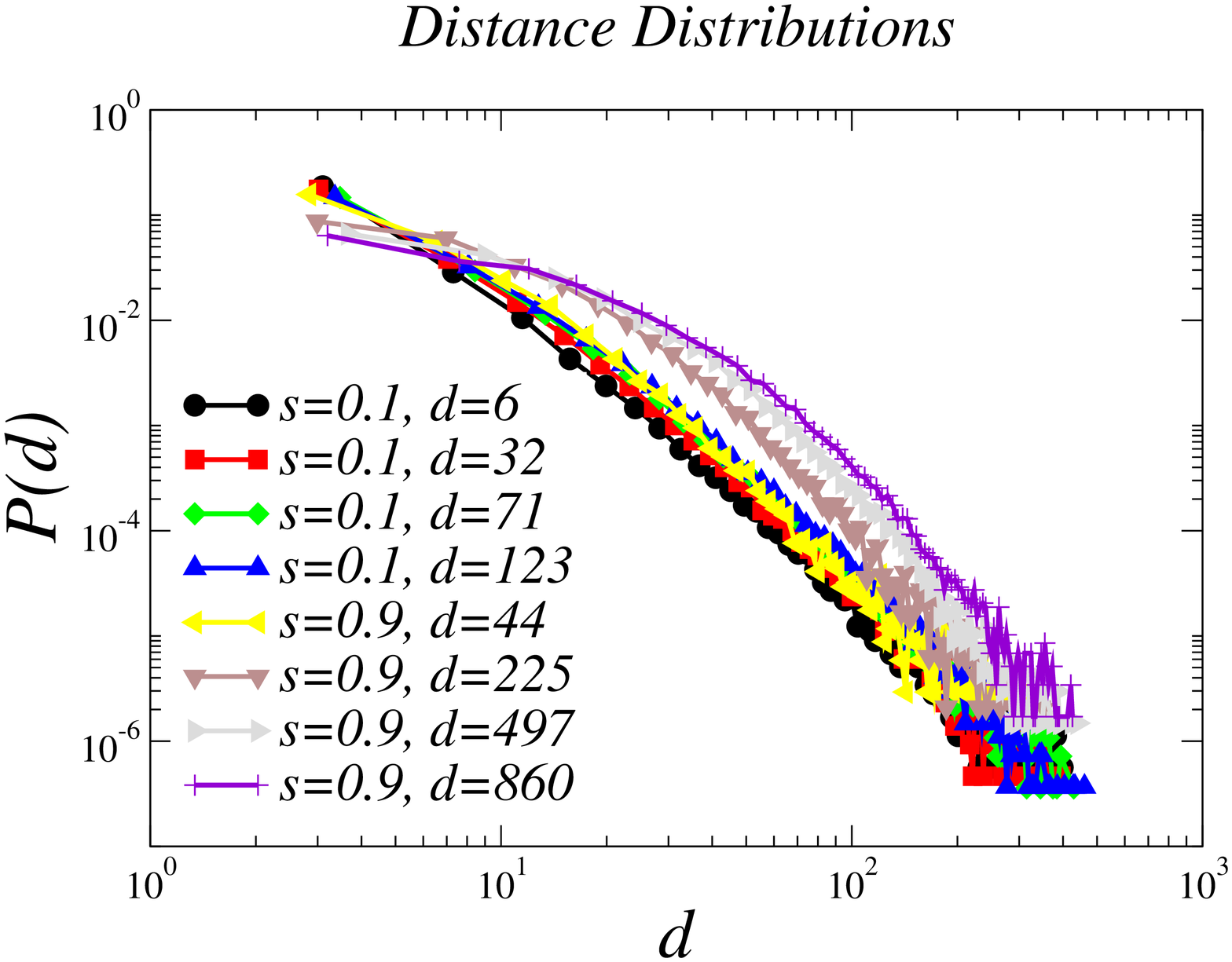}
\vspace{-1mm}
\\
\small{(a)} & \small{(b)} & \small{(c)} & \small{(d)}
%\small{(a) Degree distribution with} & \small{(b)
%Degree distribution with} & \small{(c) Indegree-Outdegree} &
%\small{(d) Distance distribution}
%\\
%\small{undirected network links} & \small{directed network links} &
%\small{correlation} & \small{}
\end{tabular}
\end{center}
\vspace{-6mm} \caption{Behavior of degree and distance distributions
for network models of a CVD thin film growth.}
\label{fig:network-models} \vspace{-6mm}
\end{figure*}

In order to explore existence of such a network behavior during thin
film and columnar nanostructure growth, we developed 3D Monte Carlo
simulations that take into shadowing, re-emission, surface
diffusion, and noise effects. These effects simulate the evolution
of surface topography and also the simulation environment allows us
to record the trajectories of re-emitted atoms. As an example,
Figure \ref{fig:fig_7} shows the snapshot top view images of two
surfaces simulated for a CVD type of deposition, at two different
sticking coefficients. Figure \ref{fig:fig_7} also displays their
corresponding particle trajectories projected on the lateral plane.
Qualitative network behavior can easily be realized in these
simulated morphologies as the trajectories of re-emitted atoms
``link'' various surface points. It can also be seen that larger
sticking coefficients (Figure \ref{fig:fig_7}(a) and Figure
\ref{fig:fig_7}(c)) leads to fewer but longer range re-emissions,
which are mainly among the peaks of columnar structures. Therefore,
these higher surface points act as the ``nodes'' of the system. This
is due to the shadowing effect where initial particles
preferentially head on hills. They also have less chance to arrive
down to valleys because of the high sticking probabilities (see also
particle A illustrated in Figure \ref{fig:fig_3}). On the other
hand, at lower sticking coefficients (Figure \ref{fig:fig_7}(b) and
Figure \ref{fig:fig_7}(d)), particles now go through multiple
re-emissions and can link many more surface points including the
valleys that normally shadowed by higher surface points (e.g.
particle B in Figure \ref{fig:fig_3}).

Another interesting observation revealed in our Monte Carlo
simulations was the dynamic change of network behavior on the
trajectories of re-emitted particles. Figure \ref{fig:fig_8} shows
top view images and their corresponding particle trajectories
obtained from the simulations for a sticking coefficient of $s =
0.9$, but this time at different film thicknesses that is
proportional to the growth time. The dynamic change in the network
topography can be clearly seen: at initial times, when the hills are
smaller and more closely spaced, the re-emitted particles travel
from one hill to another one or to a valley. However, as the film
gets thicker, and some hills become higher than the others and get
more separated, particles travel longer ranges typically among these
growing hills. The shorter hills that get shadowed become the
valleys of the system. It is expected that this dynamic behavior
should be strongly dependent on the values of sticking coefficients
and angular distribution of the incident flux of particles, which
determine the strength of re-emission and shadowing effects,
respectively. In other words, each deposition technique and material
system can have different dynamic network behavior that can lead to
various kinds of network systems.
%The dynamic network among the
%surface points of a mounded CVD grown film can be quite different
%than the one among the nanorod and nanospring structures formed in
%an oblique angle deposition system.

To make some initial observations on the network characteristics
based on our network-based models of the growth dynamics, we plotted
the degree and distance distributions in
Figure~\ref{fig:network-models} for a thin film of size
$512\times512$ lattice units. We used each lattice unit on the thin
film as a node in the corresponding network model and each
re-emission as a directed/undirected link between the nodes of the
surface locations. We developed the network models for snapshots of
the growth where each snapshot being composed of
$10\times512\times512$ particles' trajectories. We took four
snapshots at different film thickness $d$. Since the complete growth
process is very long this many particles, in some sense, samples the
surface morphology. We did this network modeling for two different
thin film growths, one with sticking coefficient $s=0.1$ and the
other with $s=0.9$.

In this manner, Figure~\ref{fig:network-models}(a) and (b) shows the
degree distribution for the network models of the snapshots when the
links are undirected and directed respectively. Overall, the degree
distributions exhibit an exponential behavior while becoming
power-law as time progresses during the growth. This means that the
interrelationship of the surface points become more dominant and
some nodes (i.e., columnar structures) on the surface become the
main hubs. The degree distributions are quite well characterized
even though the growth dynamics are very chaotic. Another
interesting observation is that, as time progresses, the degree
distribution for the case with high $s$ converges to the one with
low $s$, which is a non-intuitive result.

Figure~\ref{fig:network-models}(c) shows the relationship between
the indegree and outdegree by plotting the average outdegree of
nodes with a particular indegree value. From this graph also, it
seems that the degree distributions converge to a common behavior as
time progresses even though sticking coefficients are quire
different. Similarly, Figure~\ref{fig:network-models}(d) shows the
distance distribution of the links in the network, which clearly
exhibits a power-law structure. The network model, again, clearly
captures the behavior and shows that a higher sticking coefficient
yields larger average distance with a pseudo-power-law structure.

%Using the Monte Carlo simulation and grid network model approaches
%explained above, we will investigate various dynamic network behaviors,
%including, but not limited to:
%(i) Degree distribution of network topologies constructed by re-emissions
%of particles,\\
%(ii)    Distribution of point-to-point traffic rates of particles
%(e.g. atoms or molecules),\\
%(iii)   Inter-arrival time (i.e. inter-arrival distance/speed) distribution
%of particles in a traffic stream from a point to another point on the film,\\
%(iv)    Inter-arrival time distribution of particles arriving at a particular
%point either as the initial falling point, as a re-emission point, or
%as the final point,\\
%(v) Routing of atoms (e.g., do all atoms go to their final destination
% by means of the shortest path or other factors involved?) from their
%initial point of falling on the film to their final point of settlement
%on the film,\\
%(vi)    Dynamical change in the morphology-dependent network behavior
%due to the dynamic evolution of surface topography, shadowing locations,
%and re-emission paths as briefly discussed in the previous sections, and\\
%(vii)   Effects of sticking coefficient (i.e. re-emission), oblique
%angle (shadowing), surface diffusion, and initial surface topography
%on the dynamic networking behavior during growth.\\

\vspace{-3mm}
\section{Conclusions} \label{sec:conclusion}
\vspace{-1mm}

Our initial results on the observation of dynamic network behavior
in simulated CVD thin films are very promising and indicate that a
novel network modeling approach can be developed for various
deposition systems. We showed that particles with non-unity sticking
probabilities that are re-emitted and deposited to other parts of
the surface can form a network structure constructed by the links
among each impact point, which defines nodes of the network. In
addition, due to the shadowing effect where obliquely incident
particles hit preferentially to the higher surface points, hills of
the morphology act as the hubs of the network where most of the
particles are re-emitted from these regions. Columnar morphologies
formed under high sticking coefficients promote the creation of
long-distance network links mainly among the hills, while smoother
morphologies of smaller sticking coefficient depositions leads to
the formation of shorter range but well-connected links all over the
surface points also including valleys. Therefore, this dynamic
network behavior during thin film growth strongly depends on the
sticking probabilities, presence of obliquely incident particles,
and time-dependent morphology of the growing thin film, which leads
to the realization of a rich dynamic network system. We believe that
this work can lead to an unprecedented understanding of thin film
and nanostructure growth, which has been long sought by the
researchers. However, in order to fully develop our network concept
as a viable modeling approach, more in-depth investigations are
necessary.

%\section{Acknowledgments}
%H.G. is supported by the U.S. Department of Energy under Contract
%No. DE-AC52-06NA25396.

\vspace{-3mm}
\bibliographystyle{abbrv}
\bibliography{nanonets}

\end{document}